\documentclass[runningheads]{llncs}

\usepackage{cite}
\usepackage{amsmath,amssymb,amsfonts}
\usepackage{algorithmicx}
\usepackage{algpseudocode}
\usepackage{algorithm}    
\usepackage{wrapfig}

\usepackage{graphicx}
\usepackage{textcomp}
\usepackage{xcolor}
\usepackage{verbatim}

\begin{document}

\title{Hierarchical Recursive Precision for Accelerating Symmetric Linear Solves on MXUs}
\titlerunning{Hierarchical Recursive Precision for POTRF on MXUs}%



\author{
Vicki Carrica\inst{1}\orcidID{0009-0003-8196-9787} \and
Rabab Alomairy\inst{1,2}\orcidID{0000-0001-9911-6094} \and
Evelyne Ringoot\inst{1}\orcidID{0000-0003-4661-1483} \and
Alan Edelman\inst{1}\orcidID{0000-0001-7676-3133}
}

\authorrunning{V. Carrica et al.}

\institute{
CS and AI Laboratories, Massachusetts Institute of Technology, Cambridge, MA, USA \\
\email{\{vickicar,rabab.alomairy,eringoot,edelman\}@mit.edu}
\and
Computer, Electrical and Mathematical Sciences and Engineering Division, \\
King Abdullah University of Science and Technology (KAUST), KSA \\
\email{rabab.omairy@kaust.edu.sa}
}
\maketitle

\begin{abstract}

Symmetric positive-definite system solvers based on Cholesky factorization are fundamental to many scientific applications, such as climate modeling. We present a portable, nested recursive mixed-precision solver designed for Matrix Processing Units (MXUs), including NVIDIA Tensor Cores (H200) and AMD Matrix Cores (MI300X), that assigns low-precision FP16 arithmetic to large off-diagonal blocks, while preserving high precision on diagonal blocks to ensure numerical stability. The solver is implemented in Julia, providing a high-level, hardware-agnostic interface. We demonstrate up to a 5.07$\times$ speedup relative to the diagonal-precision vendor baseline, with 100$\times$ better accuracy than pure half precision on H200, providing higher accuracy than low-precision at higher speed than high-precision. Positive performance trends are also observed on MI300X, demonstrating broad applicability across GPUs.

\end{abstract}

\section{Introduction}
Symmetric linear systems are a critical performance bottleneck in large simulations, for example, computational fluid dynamics\cite{piscaglia2023gpu}, climate modeling\cite{alomairy2025sustainably}, and Gaussian process regression\cite{gardner2018gpytorch}. Direct solution methods of such symmetric linear systems involve calculating their Cholesky decomposition, which relies on triangular solve (TRSM) and symmetric rank-k update (SYRK). 
These systems have traditionally been solved in uniform high precision, creating a computational bottleneck on modern AI accelerators. While modern lower-precision Matrix Processing Units (MXUs), such as NVIDIA Tensor Cores and AMD Matrix Cores ~\cite{carrica2025accelerating}, deliver high throughput for low precision operations, Cholesky requires numerical stability to prevent divergence. Because scientific computing demands high-accuracy results, standard uniform-precision solvers cannot safely exploit these fast MXUs.
Deploying these hardware capabilities in scientific computing therefore requires a mixed-precision approach where high accuracy and improved speed are combined.
In this work, we demonstrate that integrating mixed-precision into the well-known multi-stage recursive subdivision algorithm enables leveraging the capabilities of MXUs and results in significant speed-ups while maintaining higher accuracy than low-precision solutions.
\textit{Recursive subdivision}~\cite{eliahu2015frpa, andersen2001recursive, bosilca2014power} is a well-known strategy to enhance data locality and parallelism in Cholesky factorization through introducing matrix-multiplies.
While prior recursive formulations typically apply only to the diagonal POTRF operation—leaving off-diagonal updates (TRSM and SYRK) in standard blocked form—we extend this into a fully \textit{nested recursive formulation}. In our approach, the decomposition tree recursively subdivides not only POTRF but also TRSM and SYRK, 
further increasing the granularity of GEMM operations
~\cite{carrica2025toward}.
We integrate this in a \textit{tree-structured mixed-precision hierarchy} that assigns low precision (FP16) to large off-diagonal blocks and progressively higher precision (FP32 or FP64) to recursively refined diagonal regions.
This mixed-precision strategy aligns better with the recursive subdivision algorithm than
 prior mixed-precision approaches, which either focus on iterative refinement~\cite{higham2022mixed,carson2018accelerating} or apply uniform low precision to predefined matrix partitions~\cite{ltaief2024toward,zhang2025leveraging}. 
We focus in this work on diagonally dominant problems, for which this is a reasonable strategy. For non-diagonally dominant problems, other mixed-precision strategies need to be considered.
The solver is implemented as a layered mixed-precision scheme in Julia. 


The main contribution of this work is the integration of a tree-based mixed-precision scheme into a multi-stage recursive subdivision strategy in a unified portable data structure, demonstrating significant speedups of mixed-precision approaches over high-precision vendor libraries on a single GPU, while maintaining higher accuracy than low precision.
On NVIDIA H200 GPUs, our solver achieves a $5.32\times$ speedup over FP64 cuSOLVER, and a 5.07$\times$ speedup against cuSOLVER FP32, with $100\times$ better accuracy than pure FP16 and 88\% of Pure FP16's peak throughput. Similar trends are observed on AMD MI300X, showcasing the cross-platform scalability and practicality of our approach.

\section{Related Work}
\label{sec:background}

\subsection{Recursive Decomposition}
Recursive formulations have long been studied as a strategy to improve data locality and expose parallelism in dense linear algebra. 
Dense linear algebra algorithms such as the Cholesky decomposition are often bounded by communication cost, rather than arithmetic cost, motivating the development of communication-avoiding Cholesky implementations \cite{ballard2009communication}. 
Early efforts in dense matrix settings—such as~\cite{andersen2001recursive, eliahu2015frpa}—proposed recursive blocked factorizations, and 
recursive formulations for hierarchical low-rank solvers~\cite{grasedyck2009domain, chen2022solving}, 
where recursion is limited to the diagonal factorization, with off-diagonal updates computed via standard blocked methods following LAPACK-style parallelism~\cite{anderson1999lapack}. 
More recently, optimizing task granularity in hybrid CPU-GPU environments has been done through recursive task graphs~\cite{faverge2023programming, furmento2025optimizing, ren2025accelerating}.
For GPU hardware,
Charara et al.~\cite{charara2017framework}
 introduced a recursive triangular solver (TRSM) designed to exploit data locality and task decomposition on GPU architectures, which Carrica et al.~\cite{carrica2025toward} implemented as a recursive GPU TRSM in Julia, demonstrating its benefits on tensor-core hardware. 
 While existing approaches thus leverage standard blocked updates \cite{charara2019batched}, to the best of our knowledge, they typically do not extend the recursive approach across all three phases  —POTRF, TRSM, and SYRK. Our approach evaluates a \textit{recursive GPU-based SYRK} extending across all three phases, along with a mixed-precision variant that strategically exploits low-precision GEMM acceleration, demonstrating it results in performance benefits over non-nested approaches, hereby addressing the linear solve bottleneck in scientific computing by accelerating this phase.
 

\subsection{Mixed-Precision Linear Algebra}
Mixed-precision techniques have gained significant attention as a means to accelerate numerical algorithms while reducing energy consumption. Higham et al.~\cite{higham2022mixed},  along with others~\cite{buttari2007mixed,baboulin2009accelerating,carson2018accelerating,haidar2018harnessing,carson2017new,alomairy2025scalable} studied mixed-precision iterative refinement for dense and sparse systems. More recently, task-based and batched GPU solvers have explored integrating low-precision arithmetic at the kernel level~\cite{zhang2025leveraging,alomairy2022high,ltaief2024toward,alomairy2025sustainably}, often relying on uniform block-level precision assignment. Other approaches apply precision dynamically based on sub-matrix norms~\cite{alomairy2025sustainably} or use non-uniform thresholds for block low-rank LU factorization~\cite{amestoy2023mixed}.
These strategies are not directly aligned with the algorithmic recursion of the Cholesky factorization. In our approach for diagonally dominant matrices, FP16 precision is assigned to large off-diagonal blocks and higher precision (FP32 or FP64) to recursively refined diagonal region, aligning a \textit{tree-structured precision hierarchy} with the recursive subdivision strategy.
This enables controlled use of MXU-accelerated low-precision GEMM operations without compromising stability. Our approach also integrates lightweight per-block quantization and dequantization~\cite{lang2024comprehensive}, ensuring robustness when operating near FP16's dynamic range limits(Section \ref{sect:quant}).
To our knowledge, this is the first work to unify recursive decomposition and layered mixed-precision execution in a hardware-portable solver framework.

\section{Methods}
\label{sec:methods}
\subsection{Julia Language abstractions}
The implementation of a recursive mixed-precision Cholesky solver is facilitated by the Julia programming language~\cite{bezanson2017julia}. 
Julia’s multiple dispatch and type inference with compile-time specialization permits writing single hardware-agnostic calls to specialized core operations—\texttt{POTRF}, \texttt{TRSM}, \texttt{SYRK}, and \texttt{GEMM}—which the compiler then resolves and optimizes automatically based on data type (e.g., \texttt{FP16}, \texttt{FP32}, \texttt{FP64}), enabling a unified mixed-precision recursive solver framework,
combining portability and high performance~\cite{ringoot2025gpu,giordano2022productivity,alomairy2024dynamic}.
In previous work, a full-cycle low-level implementation of \texttt{TRSM} and \texttt{TRMM}, directly compiled to GPU machine language, was demonstrated to provide performance on par with vendor-optimized functions\cite{carrica2025toward}. In the current work, we focus on the recursive implementation of Cholesky, and lean on low-level vendor library implementation of \texttt{POTRF}, \texttt{TRSM}, \texttt{SYRK}, and \texttt{GEMM} through multiple dispatch for the base cases. This algorithm could in the future be a critical building block towards the performant low-level hardware-agnostic implementation of Cholesky decomposition for small and large data sizes for all GPU hardware.



\subsection{Nested Recursive Algorithm Design}

We implement a nested recursive Cholesky algorithm that decomposes symmetric positive-definite (SPD) matrices by hierarchically breaking down all three core computational phases: POTRF (Cholesky factorization), TRSM (triangular solve), and SYRK (symmetric rank-$k$ update). By replacing standard TRSM and SYRK updates with their recursive counterparts, we expose more parallelism and achieve deeper task granularity. This structure increases arithmetic intensity and enables greater utilization of BLAS-3 routines, particularly GEMM, which are highly optimized on modern GPU architectures with MXUs. The nested recursive structure allows for greater memory hierarchy utilization. By recursively dividing the problem until blocks fit within the L2 cache, it enhances data locality and maximizes cache reuse compared to standard flat algorithms.



\begin{figure}[h]

    \centering
    \includegraphics[width=0.7\textwidth]{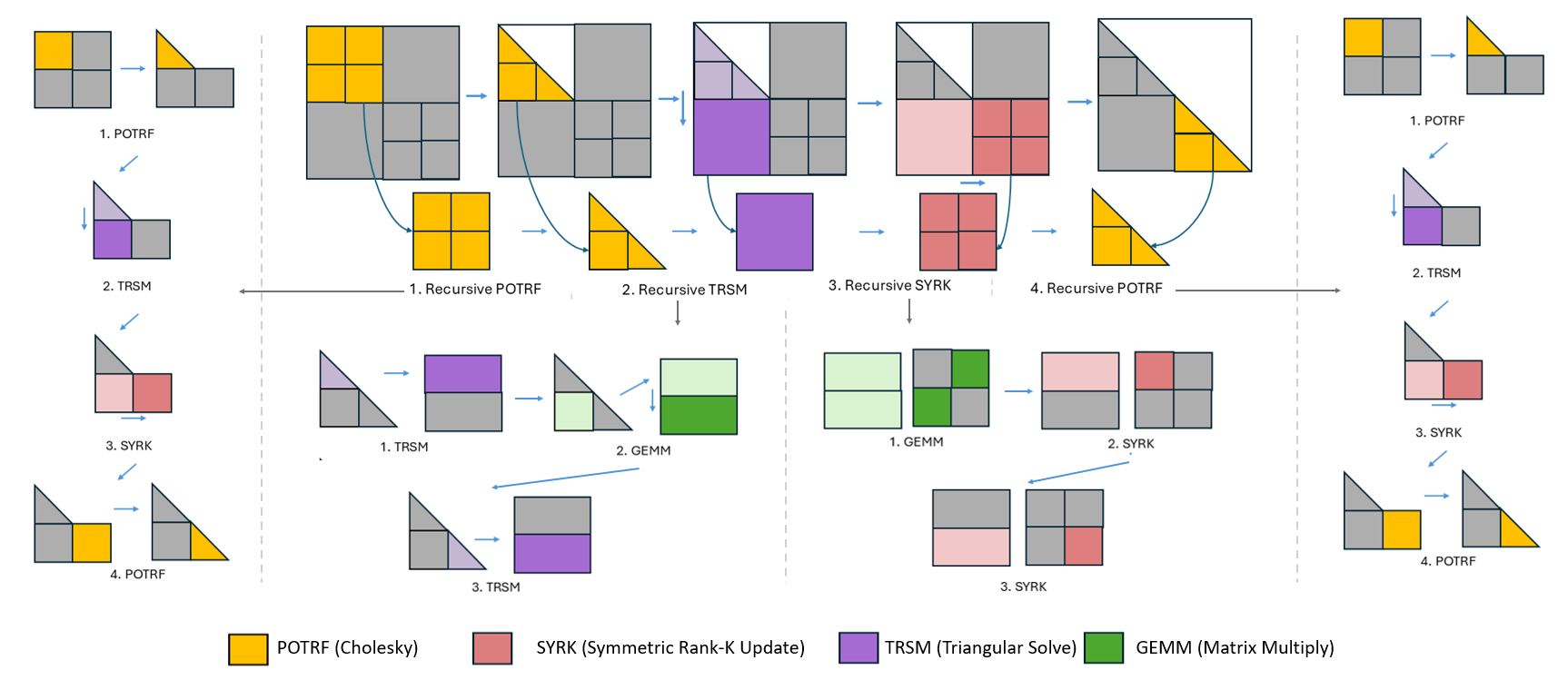}
    \caption{Nested Recursive Cholesky Decomposition Tree}
    \label{fig:decomposition_tree}
    \vspace{-15pt}
    \end{figure}

For TRSM, this form of recursion was first introduced for GPUs in~\cite{charara2019batched}, and its portable Julia implementation was recently explored in~\cite{carrica2025toward}. 
\vspace{-15pt}
\begin{algorithm}[H]
\caption{Tree-POTRF$_L(A,b)$: Nested-recursive Cholesky (lower)}
\label{alg:tree-potrf}
\scriptsize
\begin{algorithmic}[1]
\Require $A\in\mathbb{R}^{n\times n}$ SPD; leaf size $b$
\If{$n\le b$} \State $A \gets \mathrm{POTRF}_L(A)$ 
\State \Return \EndIf
\State Choose $n_1$ (e.g.\ $n_1=\lfloor n/2\rfloor$); $n_2=n-n_1$
\State Partition $A=\left[\begin{smallmatrix}A_{11}&0\\A_{21}&A_{22}\end{smallmatrix}\right]$
\State \Call{Tree-POTRF$_L$}{$A_{11},\,b$} \Comment{$A_{11}\gets L_{11}L_{11}^\top$}
\State \Call{Tree-TRSM$_{R,L,T}$}{$A_{21},\,L_{11},\,b$} \Comment{$A_{21}\gets A_{21}L_{11}^{-\top}$}
\State \Call{Tree-SYRK$_L$}{$A_{22},\,A_{21},\,\alpha=-1,\,\beta=1,\,b$} \Comment{$A_{22}\gets A_{22}-A_{21}A_{21}^\top$}
\State \Call{Tree-POTRF$_L$}{$A_{22},\,b$}
\end{algorithmic}
\end{algorithm}
\vspace{-15pt}
Algorithm~\ref{alg:tree-potrf} outlines the recursive Cholesky factorization for an SPD matrix $A$ (see Figure~\ref{fig:decomposition_tree}). At each level, the matrix is split into a diagonal block $A_{11}$ and a trailing submatrix $A_{22}$. The decomposition proceeds through four recursive stages: (1) diagonal POTRF on $A_{11}$ (line 7), (2) triangular solve (TRSM) on the off-diagonal block (line 8), (3) symmetric rank-$k$ update (SYRK) on the trailing submatrix $A_{22}$ (line 9), and (4) recursive POTRF on the updated $A_{22}$ (line 10).
Algorithm~\ref{alg:tree-trsm} applies TRSM recursively to solve $B \gets B L^{-T}$ for a lower-triangular matrix $L$. If the matrix size falls below the threshold $b$, a standard TRSM kernel is used (line 2). Otherwise, $L$ and $B$ are split, and the update $B_2 \gets B_2 - B_1 L_{21}^T$ is computed using GEMM (line 7). Similarly, Algorithm~\ref{alg:tree-syrk} performs the SYRK update $C \gets \beta C + \alpha A A^T$ recursively. The off-diagonal contribution $C_{21} \gets \beta C_{21} + \alpha A_2 A_1^T$ is explicitly computed via GEMM (line 7), while the diagonal blocks are handled recursively.


\begin{figure}[t]
\small
\begin{minipage}[t]{0.49\textwidth}
\begin{algorithm}[H]
\caption{\small Tree-TRSM$_{R,L,T}(B,L,b)$} 
\label{alg:tree-trsm}
\begin{algorithmic}[1]
\scriptsize
\Require $B\in\mathbb{R}^{m\times n}$, $L\in\mathbb{R}^{n\times n}$ lower-tri; leaf $b$
\If{$\min(m,n)\le b$} 
\State $B\gets \mathrm{TRSM}(\ldots;\,L,B)$
\State \Return 
\EndIf
\State $n_1=\lfloor n/2\rfloor$, Split $L=\left[\begin{smallmatrix}L_{11}&0\\L_{21}&L_{22}\end{smallmatrix}\right]$, $B=[\,B_1\mid B_2\,]$
\State \Call{Tree-TRSM}{$B_1,L_{11},b$}
\State $B_2 \gets B_2 - B_1 L_{21}^\top$ 
\State \Call{Tree-TRSM}{$B_2,L_{22},b$}
\end{algorithmic}
\end{algorithm}
\end{minipage}\hfill
\begin{minipage}[t]{0.49\textwidth}
\begin{algorithm}[H]
\caption{\small Tree-SYRK$_L(C,A,\alpha,\beta,b)$}
\label{alg:tree-syrk}
\begin{algorithmic}[1]
\scriptsize
\Require $C\in\mathbb{R}^{n\times n}$ sym.; $A\in\mathbb{R}^{n\times k}$; leaf $b$
\If{$n\le b$} 
\State $C\gets \mathrm{SYRK}(\ldots;\,\alpha,A,\beta,C)$
\State \Return 
\EndIf
\State $n_1=\lfloor n/2\rfloor$, Split $A=\left[\begin{smallmatrix}A_1\\A_2\end{smallmatrix}\right]$, $C=\left[\begin{smallmatrix}C_{11}&0\\C_{21}&C_{22}\end{smallmatrix}\right]$
\State \Call{Tree-SYRK}{$C_{11},A_1,\alpha,\beta,b$}
\State $C_{21} \gets \beta C_{21}+\alpha A_2 A_1^\top$ 
\State \Call{Tree-SYRK}{$C_{22},A_2,\alpha,\beta,b$}
\end{algorithmic}
\end{algorithm}
\end{minipage}
\end{figure}
\vspace{-5pt}

\subsection{Hierarchical Mixed Precision on the recursion structure}
\label{sec:mixed-precision}

This hierarchical recursion naturally supports our \textit{layered mixed-precision scheme} integrated into our recursive Cholesky algorithm. 

Specifically, we assign: (i) \textbf{Low precision (e.g., FP16)} to large, off-diagonal blocks that dominate the computation but are less sensitive to numerical error. These are typically updated using GEMM during TRSM and SYRK phases. (ii) \textbf{Higher precision (e.g., FP32 or FP64)} to diagonal blocks, which undergo POTRF and are more susceptible to round-off errors. These blocks contribute directly to the factorization's stability and accuracy.

This layered precision design (Figure~\ref{fig:precision-structure}) maximizes performance while preserving overall stability.  In the context of Cholesky, this assignment is particularly effective because the diagonally dominant matrices we target have the most numerically sensitive operations near the diagonal.

Our method is based on a custom recursive data structure implemented in Julia, which represents a symmetric matrix $A$ as a tree of submatrices. 
In terms of memory usage, this algorithm requires no additional memory during computation. The hierarchical data structure relies entirely on matrix views to partition the data, and the recursive solver updates these blocks strictly in-place. 





\subsection{Layered Quantization and Dequantization Strategy}\label{sect:quant}
\begin{figure*}[t]
    \centering
    \begin{minipage}{0.38\textwidth}
        \centering
        \includegraphics[width=\linewidth]{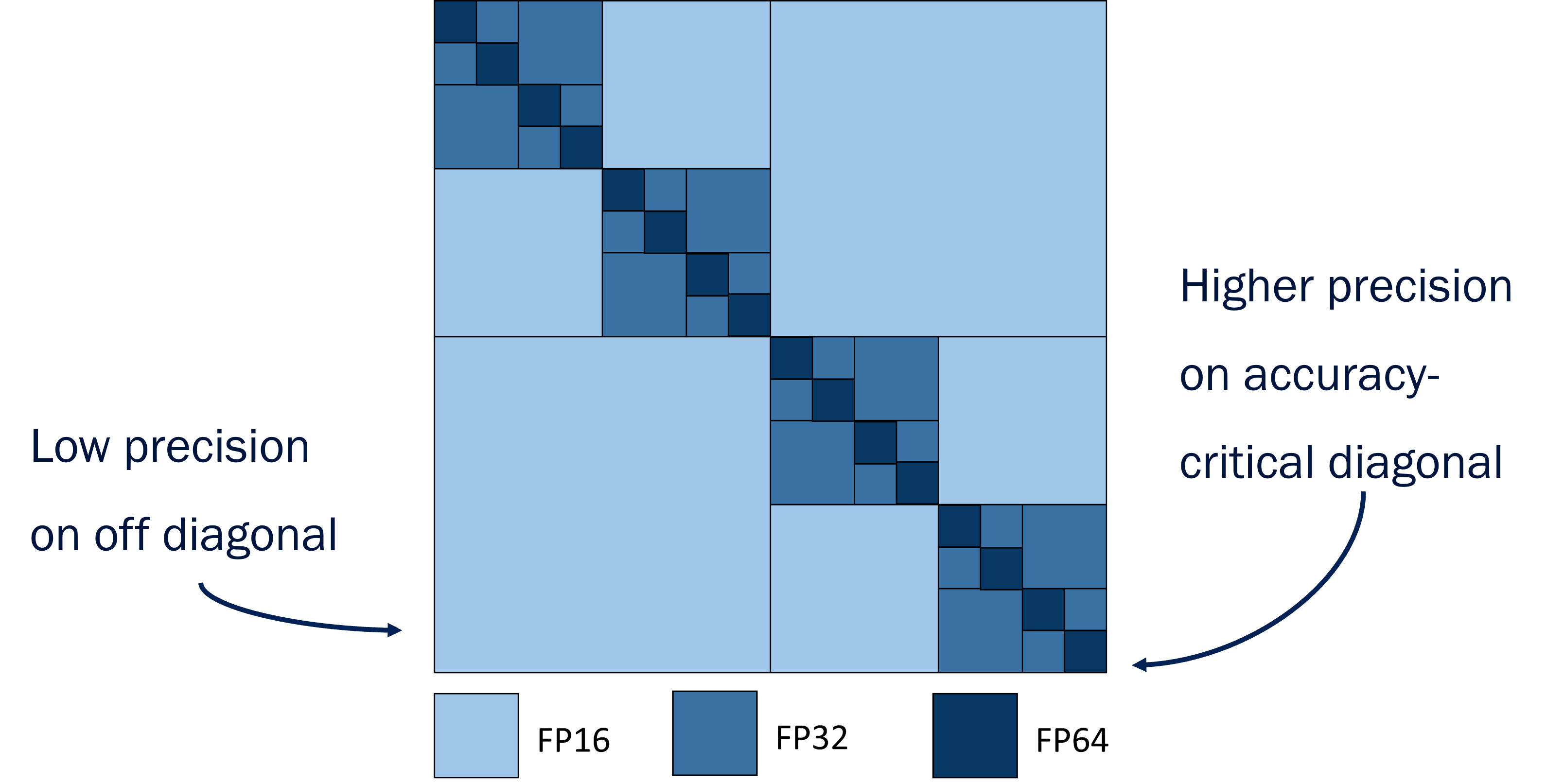}
        \caption{Recursive data structure enabling layered mixed precision.}
        \label{fig:precision-structure}
    \end{minipage}\hfill
    \begin{minipage}{0.6\textwidth}
        \centering
        \includegraphics[width=\linewidth]{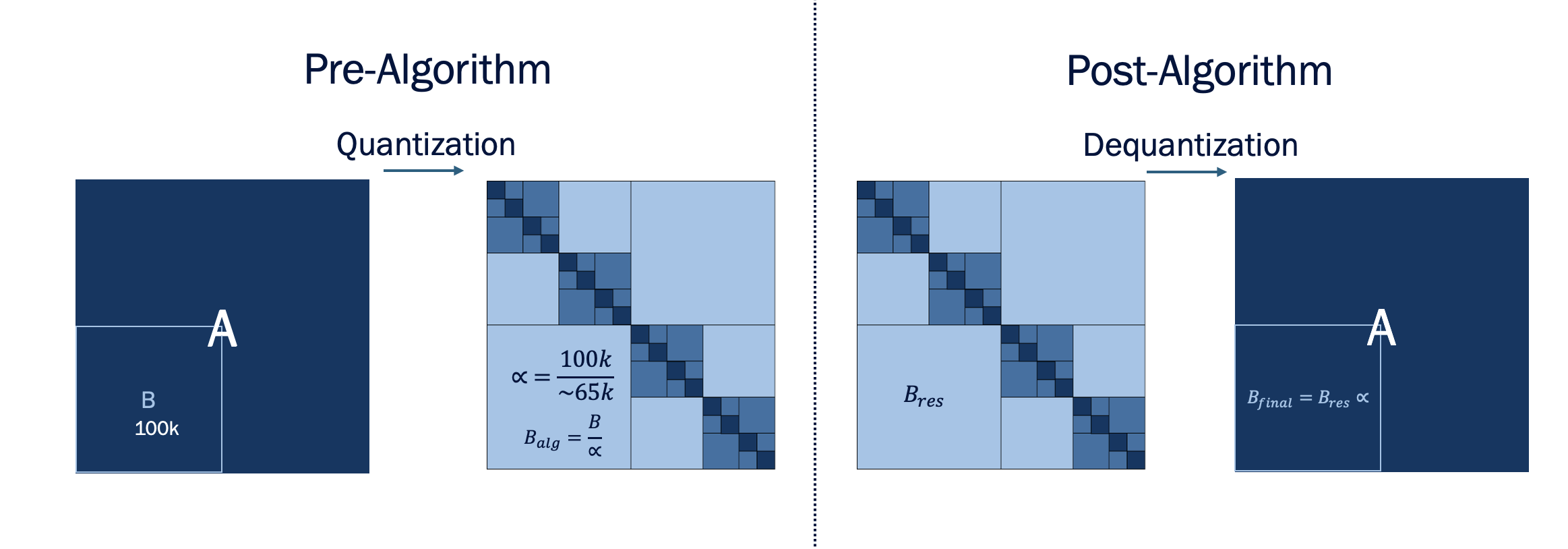}
        \caption{Blockwise quantization and dequantization workflow.}
        \label{fig:quantization}
    \end{minipage}
    \vspace{-15pt}
\end{figure*}

To mitigate FP16’s limited dynamic range and avoid numerical overflow,  we introduce a lightweight, per-block quantization and dequantization step at each level of the recursive tree. Prior to each low-precision GEMM operation, matrix blocks are rescaled into a dynamic range appropriate for FP16 execution, and then scaled back after the operation completes. 
Figure~\ref{fig:quantization} illustrates this process.

Let $A$ be the symmetric positive definite matrix to factorize, and $B$ an associated data block (e.g., right-hand side or update block) with large dynamic range. To prevent overflow when operating in low precision, we explicitly manage the dynamic range relative to $R_{\text{max}}$, the maximum representable finite value of the target format 
We compute a scaling factor $\alpha$ as
$\alpha = \max\left(1, \frac{\|B\|_\infty}{R_{\text{max}}}\right)$.

Prior to the algorithm, if the values in $B$ exceed the range (i.e., $\frac{\|B\|_\infty}{R_{\text{max}}} > 1$), this scaling compresses $B_{\text{alg}}$ into the safe range $[-R_{\text{max}}, R_{\text{max}}]$. If the values are already within range, $\alpha$ remains 1, and $B_{\text{alg}}$ preserves the original scale. Upon algorithm completion on $B_{\text{alg}}$, we obtain an intermediate result $B_{\text{res}}$, which is dequantized:
$B_{\text{final}} = B_{\text{res}} \cdot \alpha $.
This quantization--dequantization process is embedded at each level of the recursive solver and is applied only to off-diagonal blocks. Combined with our tree-structured precision layering, this enables high throughput on MXUs while preserving numerical stability. 
In practice, the quantization and dequantization of blocks take up negligible computation time.

\section{Results}

\subsection{Hardware and Software Specifications}
We evaluated our implementation on two high-end GPU platforms: the NVIDIA H200 and the AMD MI300X. In the current work, we focus on single-GPU evaluation only. The NVIDIA H200 features 141\,GB of HBM3e memory and 4.8\,TB/s of memory bandwidth. The AMD MI300X provides 192\,GB of HBM3 memory with 5.3\,TB/s of peak bandwidth, 304 compute units, and 256\,MB of L3 cache.
All benchmarks were conducted using Julia~1.12.0, with GPU support enabled via CUDA.jl (v5.9.5), AMDGPU.jl (v2.1.4), KernelAbstractions.jl (v0.9.39), and GPUArrays.jl (v11.3.1).  Our Julia-based implementation is fully portable across vendors, with performance-critical kernels dispatching to either NVIDIA’s cuBLAS/cuSOLVER or AMD’s rocBLAS/hipSOLVER backends, depending on the target device.

For our benchmarks, we generated dense, symmetric positive-definite matrices with random uniform entries. To ensure positive definiteness and robust conditioning, we added $n$ to the diagonal elements of the matrix. These well-conditioned synthetic matrices were chosen to evaluate the maximum throughput and baseline stability of our recursive mixed-precision solver. This setup isolates algorithmic performance from data-induced instabilities, while our real-world matrix evaluation (Section 4.5) assess practical applicability and limitations.

\subsection{Results Analysis on NVIDIA H200}
\label{sec:results}

We denote precision configurations as ordered lists from the outermost (largest off-diagonal) to the innermost (diagonal) recursion level. For instance, \texttt{[FP16, FP32]} assigns FP16 to the off-diagonal block and FP32 to the diagonal blocks.

\begin{figure}[htbp]
    \centering
    \begin{minipage}{0.48\textwidth}
        \centering
        \includegraphics[width=\linewidth]{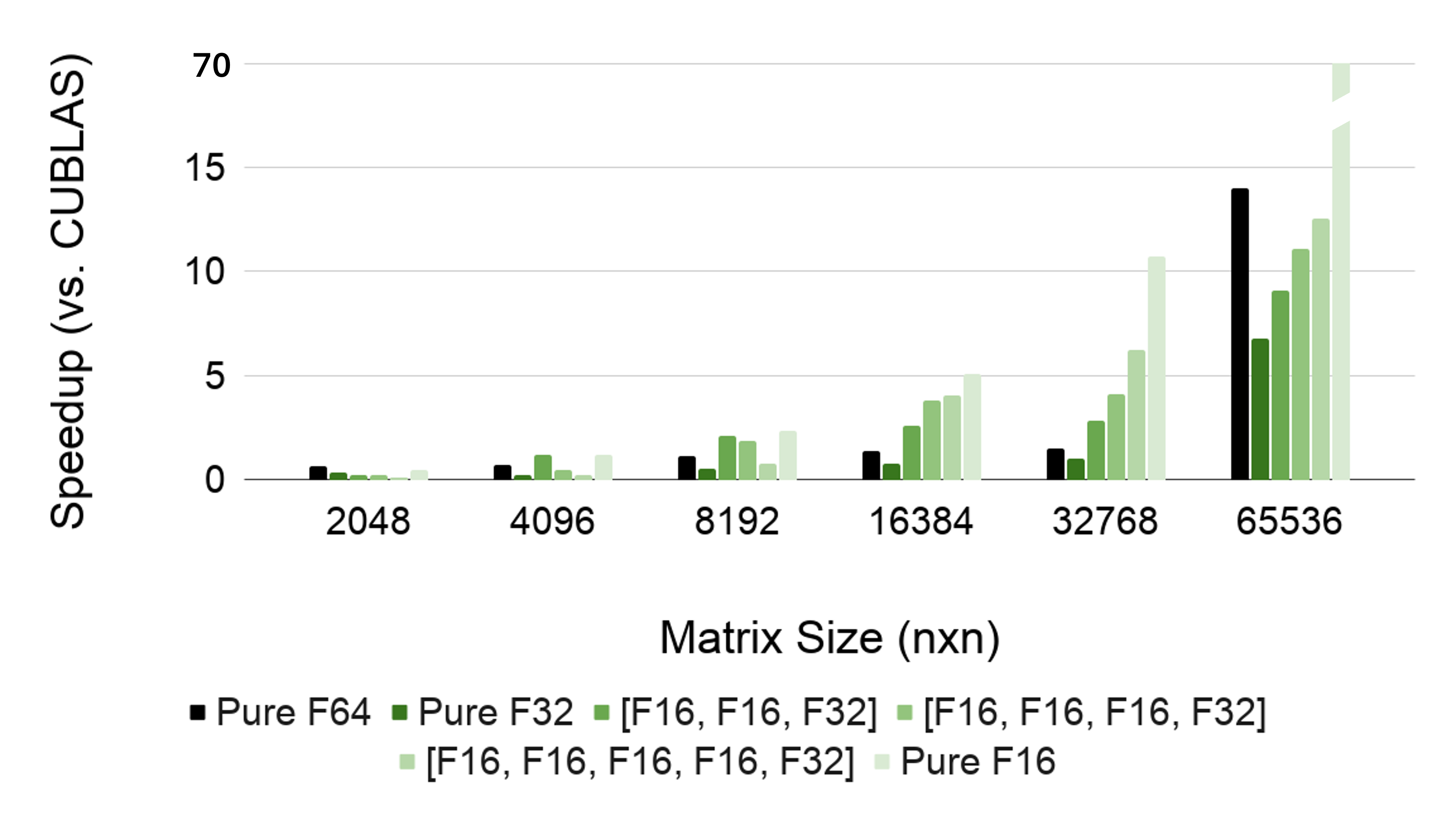}
    \end{minipage}\hfill
    \begin{minipage}{0.48\textwidth}
        \centering
        \includegraphics[width=\linewidth]{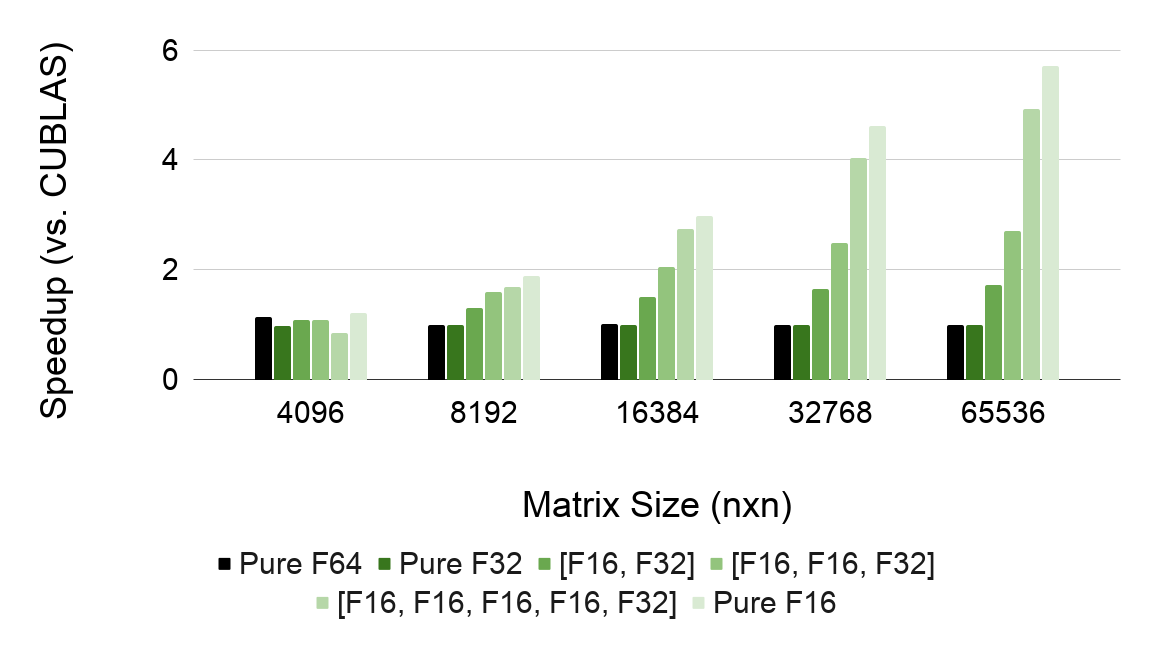}
    \end{minipage}
    \caption{Speedup of Recursive SYRK (left) and TRSM (right) on H200. Mixed-precision is normalized to diagonal precision baselines; pure FP16 uses the FP32 baseline.}
    \label{fig:syrk_trsm_h200}
\end{figure}

\paragraph{SYRK Performance.}
Figure~\ref{fig:syrk_trsm_h200} (left) reports the speedup of our Julia-based nested recursive implementation on the  H200, normalized against cuBLAS 
\texttt{SYRK} in FP64. As the matrix size increases, our recursive kernel 
consistently outperforms cuBLAS across all precision configurations. 
The FP64 variant already achieves up to a $14\times$ speedup at 
$n = 65{,}536$, benefiting from increased concurrency and improved reuse of 
tile-local data. The \texttt{Pure F16} configuration exposes the hardware’s peak throughput, reaching a 69.5$\times$ speedup relative to the FP32 baseline. By selectively exploiting this low-precision performance, mixed-precision layouts such as \texttt{[FP16, FP16, FP32]} achieve over 9$\times$ acceleration against their respective diagonal precision baseline, while the five-layer configuration \texttt{[FP16, FP16, FP16, FP16, FP32]} delivers up to a 12.5$\times$ speedup with practical accuracy.

\paragraph{TRSM Performance.}
Figure~\ref{fig:syrk_trsm_h200} (right) shows the speedup of our Julia-based recursive  implementation on the H200, normalized to 
\texttt{cuBLAS} \texttt{TRSM} in \texttt{FP64}. 
For smaller matrices, all precision configurations perform similarly to the FP64. As the matrix size increases, our recursive and mixed-precision variants 
begin to clearly outperform \texttt{cuBLAS}. The \texttt{Pure F16} implementation achieves the highest throughput, reaching up to a 5.7$\times$ speedup against the FP32 baseline at $n = 65{,}536$. Layered mixed-precision hierarchies such as \texttt{[F16, F16, F16, F16, F32]} still provide substantial gains, delivering up to 4.9$\times$ acceleration relative to their diagonal precision baseline, offering better numerical robustness than pure FP16.

\begin{figure}[t]
    \centering
    \includegraphics[width=.8\linewidth]{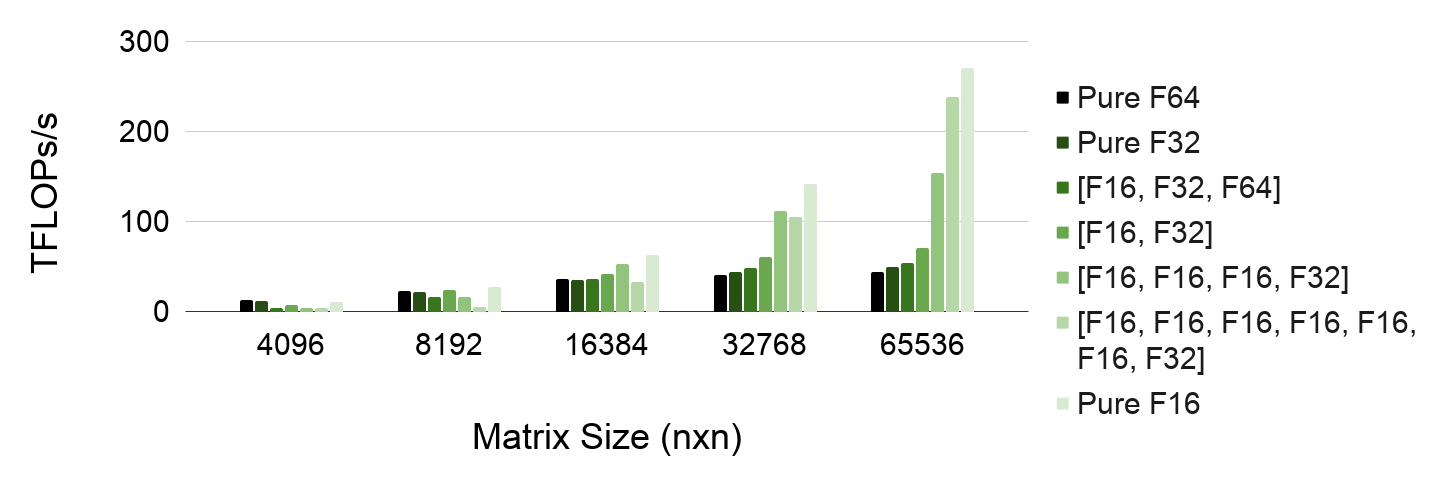}
    \caption{Cholesky Throughput on NVIDIA H200. Effective TFLOPs across matrix sizes.}
    \label{fig:cholesky_gflops_h200}
    \vspace{-12pt}
\end{figure}

\paragraph{Cholesky Throughput.}
Figure~\ref{fig:cholesky_gflops_h200} reports the Cholesky throughput on the H200.
For large problems (e.g., $n = 65{,}536$), the FP64
and FP32 recursive variants already match or slightly exceed the cuSOLVER FP64 throughput. The triple-precision \texttt{[F16, F32, F64]} configuration provides an intermediate boost, reaching roughly 54 TFLOP/s. Meanwhile, the deep mixed-precision configuration \texttt{[F16, F16, F16, F16, F16, F32]}
reaches more than $5\times$ the FP64 cuSOLVER rate. The \texttt{Pure F16} implementation
attains the highest raw throughput overall, approaching $6\times$ the FP64 baseline,
demonstrating how hierarchical mixed precision can fully exploit the H200’s MXUs for
Cholesky. At the same matrix size, both cuSOLVER FP64 and our recursive FP64
achieve roughly $65\%$ of the peak FP64 Tensor Core performance, whereas \texttt{Pure F16}
reaches about $30\%$ of the theoretical FP16 peak. This drop in efficiency at lower
precision reflects a shift in the bottleneck: as computation becomes cheaper, the
algorithm increasingly becomes limited by memory bandwidth and data movement overhead
rather than flops. The large off-diagonal FP16 GEMMs are compute-bound and the POTRF base cases and smaller block updates are memory-bound. In this regime, our deepest mixed-precision configuration
\texttt{[F16, F16, F16, F16, F16, F16, F16, F32]} sustains about $22\%$ of the FP16 peak,
yet remains roughly two orders of magnitude more accurate than \texttt{Pure F16}. 
Together, these results highlight that peak utilization alone is not the right objective:
carefully layered mixed precision can trade a modest reduction in hardware efficiency
for dramatically improved accuracy while still delivering multi$\times$ speedup over
the standard high-precision solver.

\begin{figure}[htbp]
    \centering
    \vspace{-15pt}
    \begin{minipage}{0.49\textwidth}
        \centering
        \includegraphics[width=\linewidth]{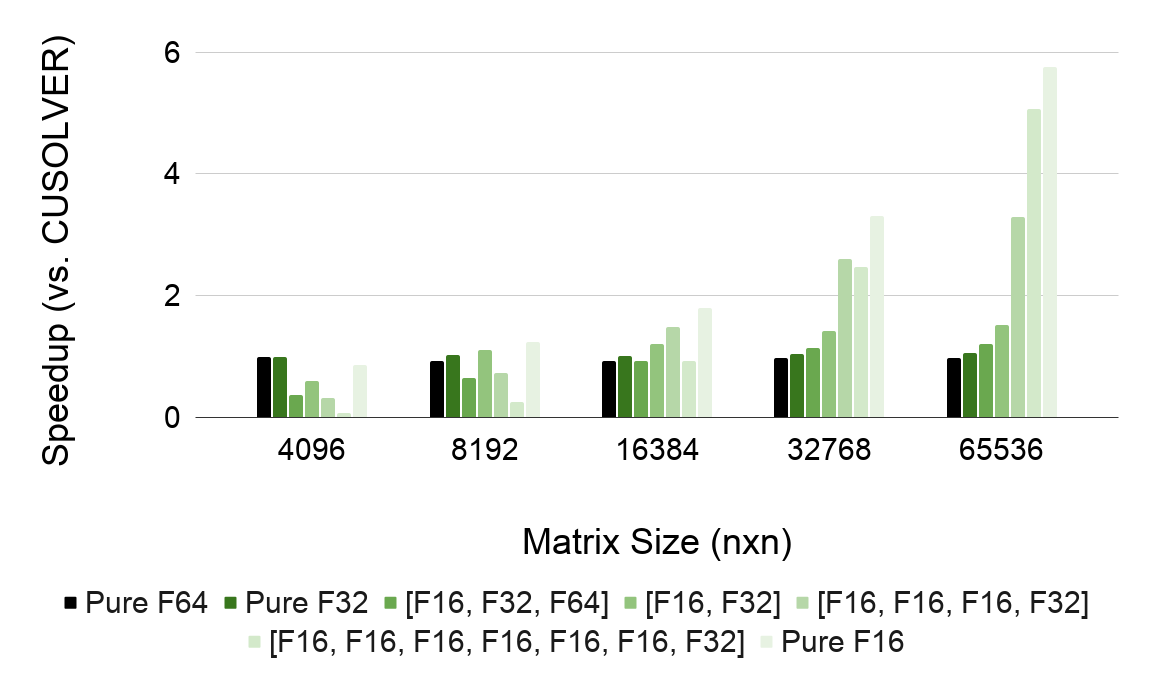}
    \end{minipage}\hfill
    \begin{minipage}{0.49\textwidth}
        \centering
        \includegraphics[width=\linewidth]{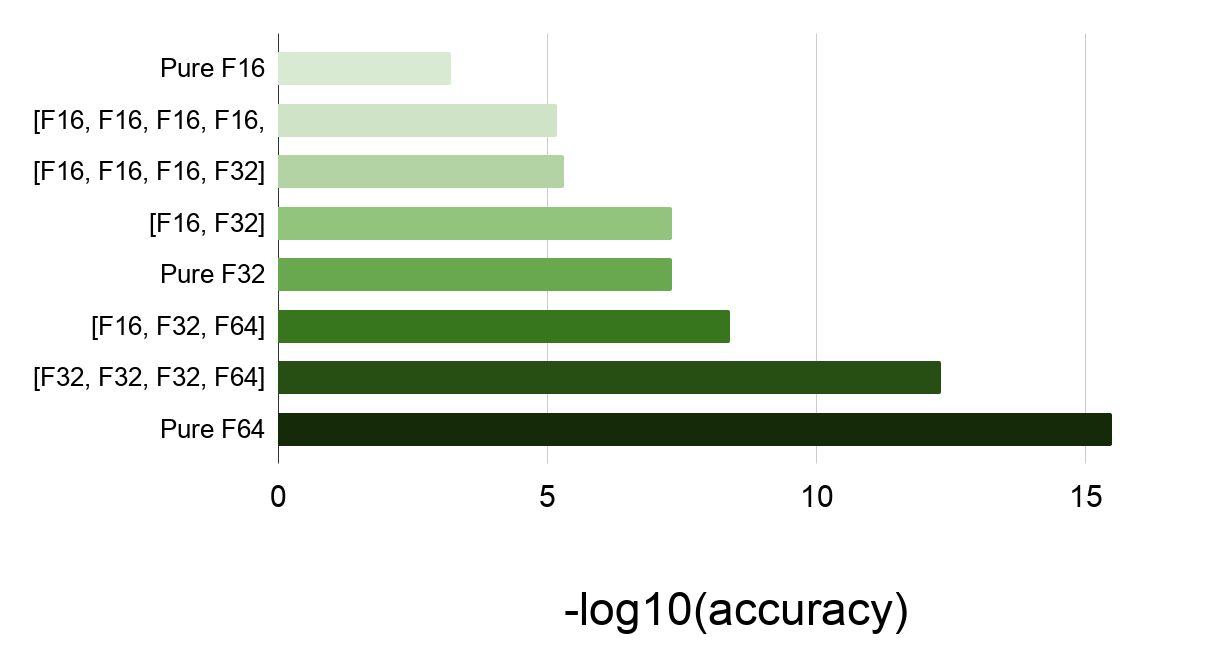}
    \end{minipage}
    \caption{Cholesky Speedup (left) and Accuracy (right) on H200. Left shows speedup normalized to diagonal precision, while right displays relative error norms ($-\log_{10}$).}
    \label{fig:cholesky_perf_acc_h200}
    \vspace{-20pt}
\end{figure}

\paragraph{Cholesky Performance.} 
Our performance evaluation of Cholesky factorization on the  H200 GPU (Fig.~\ref{fig:cholesky_perf_acc_h200}, left) highlights the benefit of precision-aware recursion. Speedups for mixed-precision configurations are normalized against the cuSOLVER baseline corresponding to their diagonal precision. Shallow configurations \texttt{[F16, F32]} yield modest speedups against cuSOLVER FP32, and the \texttt{[F16, F32, F64]} configuration overcomes initial small-matrix overhead to reach a 1.21$\times$ speedup against cuSOLVER FP64. Deeper mixed-precision hierarchies, such as \texttt{[F16, F16, F16, F32]} and \texttt{[F16, F16, F16, F16, F16, F16, F32]}, show progressively higher gains—reaching up to 3.29$\times$ and 5.07$\times$ speedups respectively relative to their FP32 diagonal for large matrices ($n = 65{,}536$). Due to the absence of a vendor FP16 solver, the deepest configuration (\texttt{Pure F16}) is evaluated against the FP32 baseline, achieving a 5.75$\times$ acceleration, highlighting the synergy between low-precision arithmetic and our recursive implementation. These gains stem from the aggressive use of fast \texttt{FP16} compute and efficient tiling strategies that exploit the H200’s high memory bandwidth and ample L2 cache.

We can break down the execution time of the full FP64 recursive solver at $n=65{,}536$, which has a runtime of 2,141\,ms. The recursive TRSM and SYRK updates account for roughly 788\,ms (37\%) and 514\,ms (24\%) of the total runtime, respectively. The remaining 839\,ms (39\%) is spent in the POTRF base cases and recursive overhead. Because the TRSM and SYRK off-diagonal updates account for over 60\% of the execution time, accelerating the performance of these specific routines via low-precision is what drives the speedups observed in large matrices.

\paragraph{Cholesky Accuracy.}
Figure~\ref{fig:cholesky_perf_acc_h200} (right) reports the accuracy of our mixed-precision Cholesky on the H200 GPU. It shows the relative error norm between the Cholesky factor computed using each mixed-precision configuration and the baseline FP64 result, plotted across matrix sizes. Note that we report the factorization error ($||A - LL^T|| / ||A||$) rather than the residual error of a solved system, as it provides a more conservative measure of numerical stability. As expected, \texttt{Pure F64} attains the highest accuracy, with more than 15 correct digits. The configuration \texttt{[F32, F32, F32, F64]} remains very close, preserving roughly 12 digits and outperforming \texttt{Pure F32}, which yields about 8 digits. Introducing FP16 at the lower levels (\texttt{[F16, F32]}) still maintains single-precision–like accuracy (around 7–8 digits), while delivering substantial speedups in our performance results. For applications requiring strict reliability, adding a top layer of FP64 (\texttt{[F16, F32, F64]}) tightly tracks the \texttt{Pure F64} baseline accuracy while still offering performance gains. Deeper hierarchies with more FP16 (\texttt{[F16, F16, F16, F32]} and \texttt{[F16, F16, F16, F16, F16, F32]}) gradually reduce accuracy to about 5–6 digits, yet remain more accurate than \texttt{Pure F16}, which drops below 4 digits. Overall, this illustrates a clear trade-off: the deepest configuration (\texttt{[F16, F16, F16, F16, F16, F16, F32]}) achieves a 5.07$\times$ speedup but retains only 5–6 digits of accuracy, whereas the \texttt{[F16, F32, F64]} configuration reaches roughly 9 digits of accuracy but with a more modest 1.21$\times$ speedup.

\subsection{Results Analysis on AMD MI300X}

\begin{wrapfigure}{r}{0.55\textwidth}
\vspace{-16pt}
    \centering
    \includegraphics[width=0.5\textwidth]{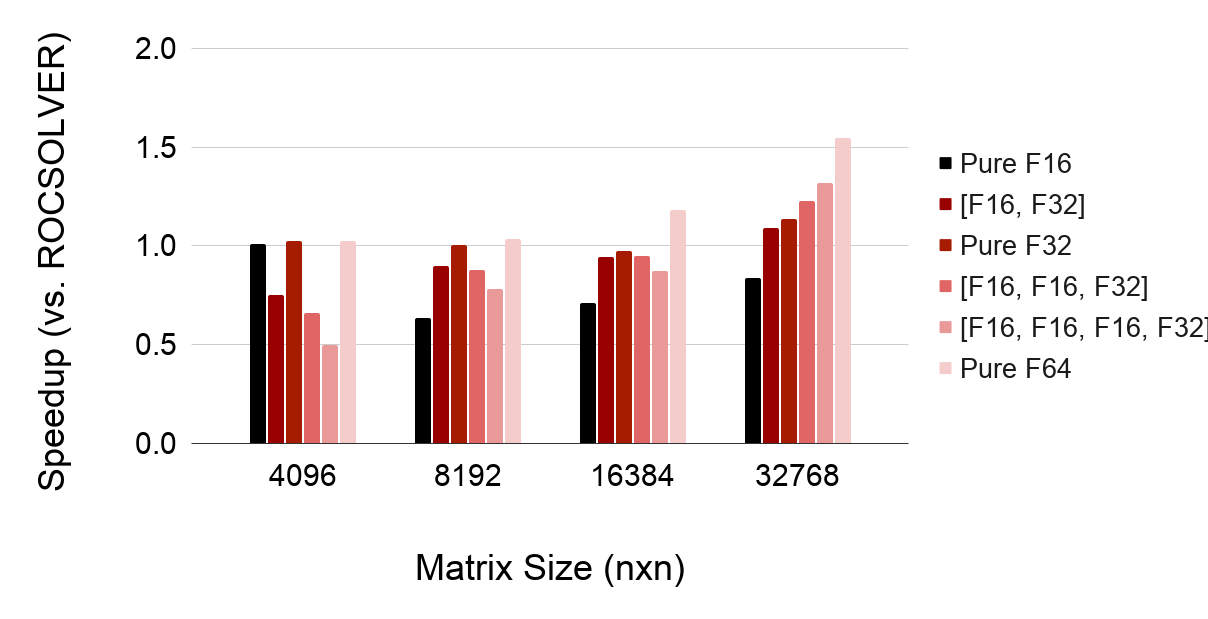}
    \caption{Cholesky Speedup on AMD MI300X. Speedups are normalized against the \texttt{rocSOLVER} baseline corresponding to their diagonal precision.}
    \label{fig:mi300x_speedup}
\vspace{-16pt}
\end{wrapfigure}

We evaluate our recursive Cholesky solver on the MI300X. To ensure a fair evaluation, speedups for mixed-precision configurations are normalized against the vendor-optimized \texttt{rocSOLVER} baseline corresponding to their diagonal precision. Figure~\ref{fig:mi300x_speedup} shows the relative speedup across matrix sizes.

Our \texttt{Pure F64} recursive solver achieves up to a 1.54$\times$ speedup over \texttt{rocSOLVER} FP64 at $n=32{,}768$, benefiting from recursive blocking and improved cache reuse on the MI300X. The \texttt{Pure F32} variant yields a 1.14$\times$ speedup over \texttt{rocSOLVER} FP32. Mixed-precision variants with increasing layering depth provide further gains relative to their diagonal baselines. For instance, the \texttt{[F16, F16, F32]} configuration reaches a 1.23$\times$ speedup against the FP32 baseline, while our deepest evaluated layout on this architecture, \texttt{[F16, F16, F16, F32]}, achieves up to a 1.32$\times$ speedup. Due to the absence of a vendor FP16 solver, the \texttt{Pure F16} configuration is evaluated against the FP32 baseline, yielding a 0.84$\times$ speedup (indicating a performance drop). 
These results confirm that our recursive, mixed-precision solver outperforms \texttt{rocSOLVER} for large matrices, though the overall gains on AMD are more modest than on NVIDIA due to the current lack of mixed-precision matrix multiplication (\texttt{GemmEx}) support in Julia AMD.


\subsection{Effect of Recursive Depth and Matrix Scale}
One of the key advantages of utilizing a nested recursive formulation is the ability to tune the performance and accuracy by changing the number of levels of recursion. The speedup of the best mixed-precision configuration scales linearly with matrix size ( Figure \ref{fig:scaling}), reaching a peak of $5.07\times$ relative to its diagonal precision baseline at $n=65{,}536$.
This trend is driven by the fact that larger matrices allow for deeper recursion hierarchies while maintaining leaf block sizes large enough to run efficiently on hardware. At these larger scales, increasing the recursive depth maximizes the proportion of floating-point operations that occur in the off-diagonal blocks, which are computed with FP16 GEMM operations. However, this strategy is less effective on smaller matrices; if the recursion is too deep relative to the matrix size, the blocks become too small, and the overhead of managing many tiny tasks outweighs the computational benefits. The exact point where MXU acceleration overtakes recursive overhead is hardware-dependent: on the H200, mixed-precision configurations become advantageous at $n = 8192$, whereas on the MI300X, the threshold is $n = 32{,}768$.

\subsection{Results on Real-World Matrices and Limitations}

\begin{figure}[t]
    \centering
    \includegraphics[width=0.8\linewidth]{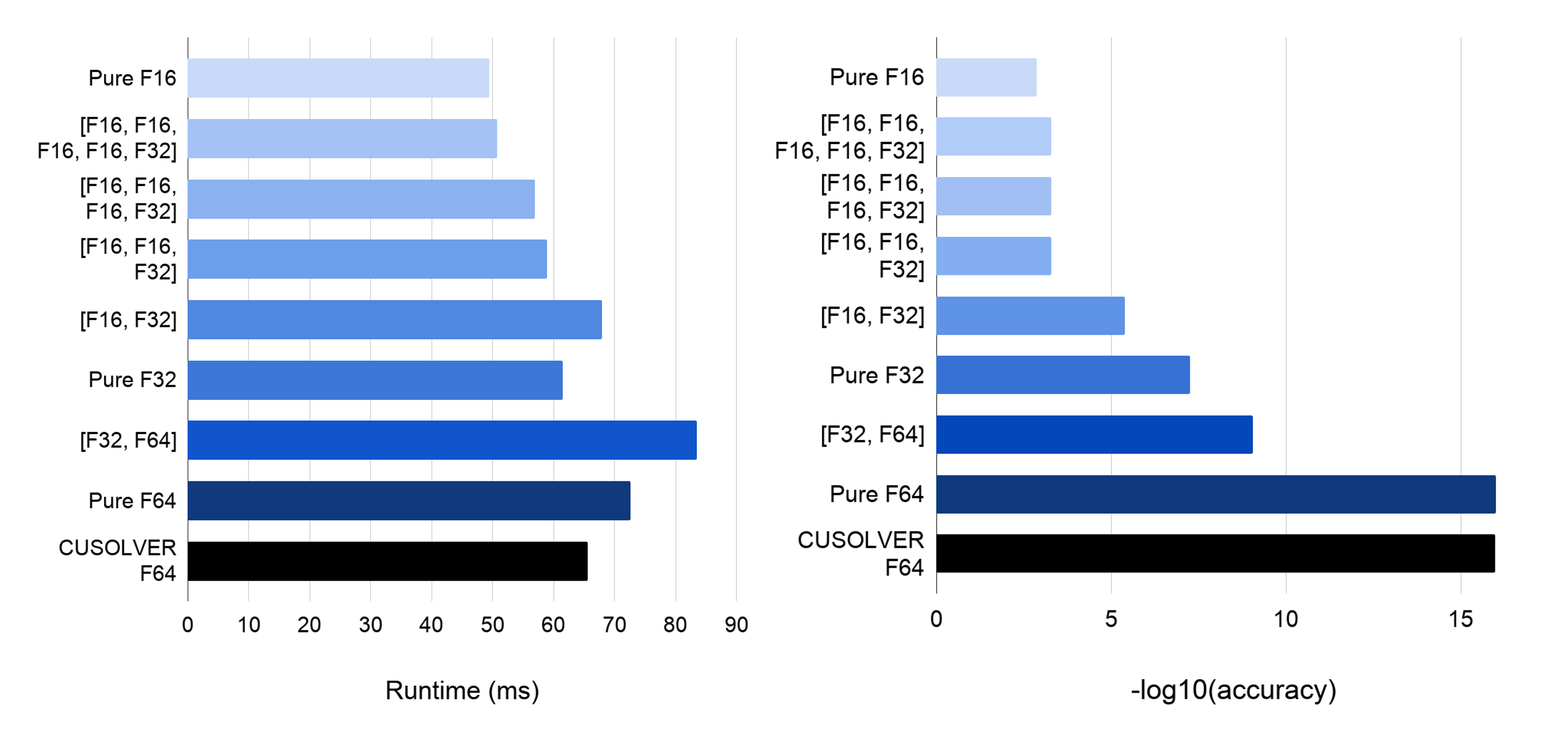}
    \caption{Runtime (left) and accuracy (right) of mixed-precision on \texttt{bodyy5} matrix.
    }
    \label{fig:bodyy5_timeacc}
\end{figure}

We evaluated our mixed-precision nested recursive Cholesky algorithm on real-world matrices using datasets from the SuiteSparse Matrix Collection, accessed via the Julia \texttt{MatrixDepot} package on an NVIDIA GPU. One key matrix we evaluated was \texttt{Pothen/bodyy5}, a sparse, symmetric positive-definite matrix derived from NASA structural engineering data ($n=18{,}589$).
As shown in Figure~\ref{fig:bodyy5_timeacc}, mixed precision retains high accuracy for moderately conditioned matrices like \texttt{bodyy5}. Our blockwise quantization effectively scales the values to prevent immediate overflow. We notice that as the precision increases from FP16 → FP32 → FP64, the runtime correspondingly increases because higher precision arithmetic requires more computational resources and memory bandwidth. Mixed-precision configurations such as [F16, F32] and [F16, F16, F32] provide an intermediate performance point, achieving substantial speedups compared to pure FP64 while maintaining greater numerical stability than pure FP16.



However, the solver is not well-suited for small matrices or matrices with a large dynamic range. For small matrices ($n < 8192$)(Figure \ref{fig:scaling})the type conversion overhead outweighs the MXU acceleration benefits, and performance is reduced. Best results are obtained for diagonally dominant matrices. Furthermore, matrices that possess extreme dynamic ranges lose positive definiteness in the process. For example, \texttt{Simon/raefsky4} ($n=19{,}779$) has a dynamic range exceeding $10^{11}$: where the quantization process compresses the lower values to zero,
destroys the positive definiteness of the sub-blocks, and causes division by zero and \texttt{NaN} propagation. Specifically, numerical errors introduced in the off-diagonal blocks can easily lead to a loss of positive definiteness in the Schur complement, even if the subsequent diagonal block is computed in high precision. Therefore, static blockwise quantization is insufficient for these cases, and future work should explore more dynamic quantization schemes.

\subsection{Portability and impact of hardware}
\begin{figure}[htbp]
    \centering
    \vspace{-15pt}
    \begin{minipage}{0.4\textwidth}
        \centering
        \includegraphics[width=\linewidth]{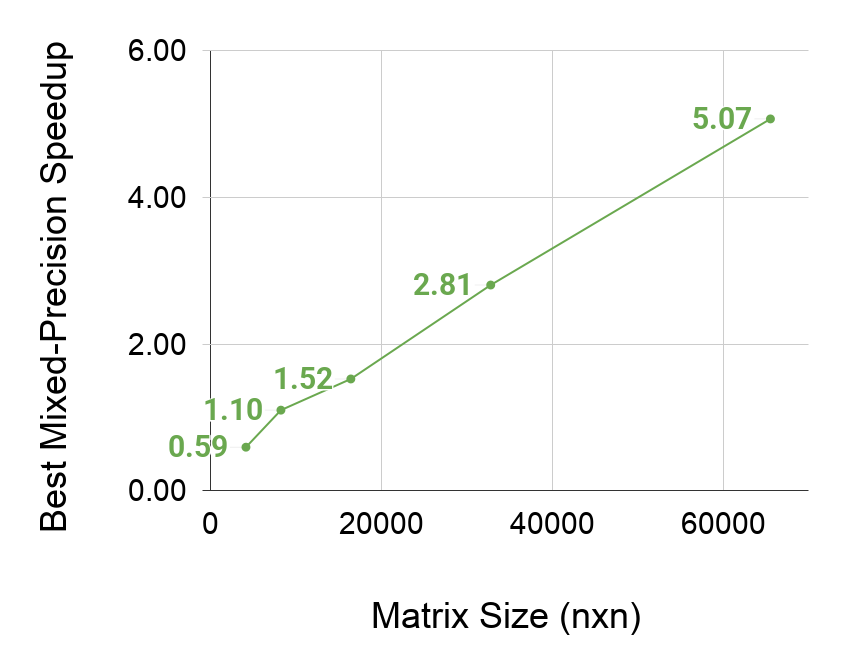}
        \caption{Speedup Scaling with Matrix Size on NVIDIA hardware. Performance gains improve as size grows, allowing deeper recursion.}
        \label{fig:scaling}
    \end{minipage}\hfill
    \begin{minipage}{0.56\textwidth}
        \centering
        \includegraphics[width=\linewidth]{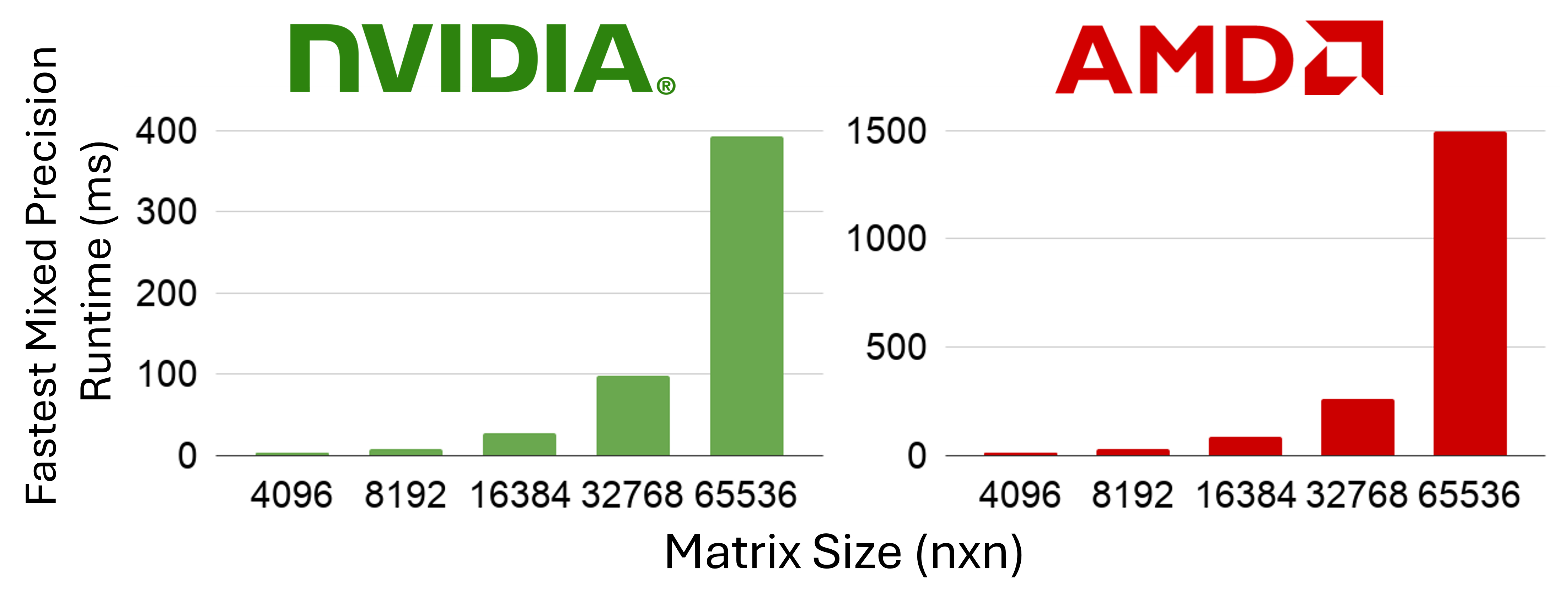}
        \caption{Cross-Platform Portability. Best mixed-precision runtime (ms) on NVIDIA H200 and AMD MI300X.}
        \label{fig:placeholder}
    \end{minipage}
    \vspace{-12pt}
\end{figure}

Figure \ref{fig:placeholder} shows the runtime of the fastest mixed-precision scenarios at each data size for the H200 and MI300X GPUs, demonstrating the portability of our implementation: with a single high-level algorithmic implementation, we cover all GPU hardware backends by using multi-dispatch to direct the base case functions to the optimized vendor library implementations. Several factors contribute to the lower performance observed on AMD compared to NVIDIA hardware, including the performance of the optimized base cases and the inclusion of mixed-precision matrix multiplication (\texttt{GemmEx}). For this benchmark, we included mixed-precision matrix multiplication for NVIDIA, but not for AMD hardware, due to its unavailability at the time in Julia. Future work into fully hardware-agnostic mixed-precision \texttt{GEMM} operations could address this and render the entire algorithm lifecycle hardware-agnostic, potentially unlocking new levels of performance not only on AMD hardware, but also on Apple GPUs.

\vspace{-12pt}
\section{Conclusion and Future Work}
\label{sec:conclusion}
We presented a portable, mixed-precision solver for symmetric linear systems based on a nested recursive Cholesky decomposition. By uniformly applying recursion across the POTRF, TRSM, and SYRK phases—including the first recursive GPU implementation of SYRK—this formulation exposes a rich hierarchy of GEMM operations that maps efficiently onto MXUs. Our Julia implementation realizes over a 5$\times$ speedup relative to diagonal-precision baselines on the  H200 and exhibits consistent improvements on the MI300X, though currently constrained by Julia's AMD \texttt{GemmEx} support.

The strategy proposed in this work is best suited for matrices that are diagonally dominant, with limited dynamic range, and are sufficiently large (e.g., $n \ge 8192$), where the computational gains of reduced-precision arithmetic outweigh the recursive overhead. Future work will explore dynamic quantization schemes and extend this strategy to broader classes of problems, such as block-sparse, banded systems, and LDL$^\top$ factorizations. 

{
\footnotesize\selectfont
\subsubsection*{Acknowledgments}
We thank the members of the Julia Lab. R.A. acknowledges the KAUST Ibn Rushd post-doctoral fellowship. The authors acknowledge the MIT ORCD, TACC at UT Austin (TG-CIS250776, allocation  CIS250776 through ACCESS,  supported by U.S. NSF 2138259, 2138286, 2138307, 2137603, and 2138296), and the AMD University Program. 
This material is based upon work supported by the US NSF (CNS-2346520, RISE-2425761, DMS-2325184, OAC-2103804, OSI-2029670), DARPA (HR00112490488), USAFR (FA8750-19-2-1000), and DoE NNSA (DE-NA0004266). The U.S. Government, its agencies and employees do not make any warranty, do not assume any liability or responsibility, do not make any endorsement for anything in this report, and do not represent that its use would not infringe privately owned rights. The views and opinions of authors expressed herein are those of the authors alone.

\subsubsection*{Disclosure of Interests}
The authors have no competing interests to declare that are relevant to the content of this article.

\bibliographystyle{splncs04}
\bibliography{sample-base}

@techreport{eliahu2015frpa,
  title={FRPA: A framework for recursive parallel algorithms},
  author={Eliahu, D. and others},
  year={2015},
  institution={EECS, UC Berkeley}
}

@inproceedings{ren2025accelerating,
  title={Accelerating Mixed-Precision Out-of-Core Cholesky Factorization with Static Task Scheduling},
  author={Ren, J. and others},
  booktitle={Proc. ISC High Perform.},
  pages={1--12},
  year={2025}
}

@inproceedings{ballard2009communication,
  title={Communication-optimal parallel and sequential {Cholesky} decomposition},
  author={Ballard, G. and others},
  booktitle={Proc. SPAA '09},
  pages={245--252},
  year={2009}
}

@article{carrica2025toward,
  title={Toward portable {GPU} performance: {Julia} recursive implementation of {TRMM} and {TRSM}},
  author={Carrica, V. and others},
  journal={arXiv preprint arXiv:2504.13821},
  year={2025}
}

@inproceedings{ltaief2024toward,
  title={Toward capturing genetic epistasis from multivariate genome-wide association studies using mixed-precision kernel ridge regression},
  author={Ltaief, H. and others},
  booktitle={Proc. SC24},
  pages={1--12},
  year={2024}
}

@article{gardner2018gpytorch,
  title={{GPyTorch}: Blackbox matrix-matrix {Gaussian} process inference with {GPU} acceleration},
  author={Gardner, J. and others},
  journal={NeurIPS},
  volume={31},
  year={2018}
}

@article{piscaglia2023gpu,
  title={{GPU} acceleration of {CFD} simulations in {OpenFOAM}},
  author={Piscaglia, F. and Ghioldi, F.},
  journal={Aerospace},
  volume={10},
  number={9},
  pages={792},
  year={2023}
}

@article{andersen2001recursive,
  title={A recursive formulation of {Cholesky} factorization of a matrix in packed storage},
  author={Andersen, B. S. and others},
  journal={ACM Trans. Math. Softw.},
  volume={27},
  number={2},
  pages={214--244},
  year={2001}
}

@article{bosilca2014power,
  title={Power profiling of {Cholesky} and {QR} factorizations on distributed memory systems},
  author={Bosilca, G. and Ltaief, H. and Dongarra, J.},
  journal={Comput. Sci. Res. Dev.},
  volume={29},
  number={2},
  year={2014}
}

@article{charara2017framework,
  title={A framework for dense triangular matrix kernels on various manycore architectures},
  author={Charara, A. and Keyes, D. and Ltaief, H.},
  journal={Concurrency Computat. Pract. Exper.},
  volume={29},
  number={15},
  pages={e4187},
  year={2017}
}

@book{anderson1999lapack,
  title={{LAPACK} Users' Guide},
  author={Anderson, E. and others},
  year={1999},
  publisher={SIAM}
}

@article{grasedyck2009domain,
  title={Domain decomposition based {H-LU} preconditioning},
  author={Grasedyck, L. and Kriemann, R. and Le Borne, S.},
  journal={Numer. Math.},
  volume={112},
  number={4},
  pages={565--600},
  year={2009}
}

@inproceedings{chen2022solving,
  title={Solving linear systems on a {GPU} with hierarchically off-diagonal low-rank approximations},
  author={Chen, C. and Martinsson, P. G.},
  booktitle={Proc. SC22},
  pages={1--15},
  year={2022}
}

@article{higham2022mixed,
  title={Mixed precision algorithms in numerical linear algebra},
  author={Higham, N. J. and Mary, T.},
  journal={Acta Numerica},
  volume={31},
  pages={347--414},
  year={2022}
}

@article{baboulin2009accelerating,
  title={Accelerating scientific computations with mixed precision algorithms},
  author={Baboulin, M. and others},
  journal={Comput. Phys. Commun.},
  volume={180},
  number={12},
  pages={2526--2533},
  year={2009}
}

@article{buttari2007mixed,
  title={Mixed precision iterative refinement techniques for the solution of dense linear systems},
  author={Buttari, A. and others},
  journal={Int. J. High Perform. Comput. Appl.},
  volume={21},
  number={4},
  pages={457--466},
  year={2007}
}

@article{carson2018accelerating,
  title={Accelerating the solution of linear systems by iterative refinement in three precisions},
  author={Carson, E. and Higham, N. J.},
  journal={SIAM J. Sci. Comput.},
  volume={40},
  number={2},
  pages={A817--A847},
  year={2018}
}

@inproceedings{haidar2018harnessing,
  title={Harnessing {GPU} tensor cores for fast {FP16} arithmetic to speed up mixed-precision iterative refinement solvers},
  author={Haidar, A. and others},
  booktitle={Proc. SC18},
  pages={603--613},
  year={2018}
}

@article{carson2017new,
  title={A new analysis of iterative refinement and its application to accurate solution of ill-conditioned sparse linear systems},
  author={Carson, E. and Higham, N. J.},
  journal={SIAM J. Sci. Comput.},
  volume={39},
  number={6},
  pages={A2834--A2856},
  year={2017}
}

@inproceedings{zhang2025leveraging,
  title={Leveraging hardware-aware computation in mixed-precision matrix multiply: A tile-centric approach},
  author={Zhang, Q. and others},
  booktitle={Proc. WAMTA},
  pages={174--185},
  year={2025}
}

@phdthesis{alomairy2022high,
  title={High-performance scientific applications using mixed precision and low-rank approximation powered by task-based runtime systems},
  author={Alomairy, R. M.},
  year={2022},
  school={KAUST}
}

@inproceedings{alomairy2025sustainably,
  title={Sustainably modeling a sustainable future climate},
  author={Alomairy, R. and others},
  booktitle={Proc. HPEC},
  pages={1--8},
  year={2025}
}

@inproceedings{lang2024comprehensive,
  title={A comprehensive study on quantization techniques for large language models},
  author={Lang, J. and others},
  booktitle={Proc. ICAIRC},
  pages={224--231},
  year={2024}
}

@misc{carrica2025accelerating,
  title={Accelerating linear solve with mixed precision nested recursive subdivision on {AI} hardware},
  author={Carrica, V. and others},
  year={2025},
  note={SC25 Poster}
}

@article{bezanson2017julia,
  title={Julia: A fresh approach to numerical computing},
  author={Bezanson, J. and others},
  journal={SIAM Review},
  volume={59},
  number={1},
  pages={65--98},
  year={2017}
}

@article{ringoot2025gpu,
  title={A {GPU}-resident memory-aware algorithm for accelerating bidiagonalization of banded matrices},
  author={Ringoot, E. and others},
  journal={arXiv preprint arXiv:2510.12705},
  year={2025}
}

@inproceedings{giordano2022productivity,
  title={Productivity meets performance: {Julia} on {A64FX}},
  author={Giordano, M. and others},
  booktitle={Proc. IEEE CLUSTER},
  pages={549--555},
  year={2022}
}

@inproceedings{alomairy2024dynamic,
  title={Dynamic task scheduling with data dependency awareness using {Julia}},
  author={Alomairy, R. and others},
  booktitle={Proc. HPEC},
  pages={1--7},
  year={2024}
}

@inproceedings{alomairy2025scalable,
  title={Scalable {Hamming} distance computation using accelerated matrix transformations},
  author={Alomairy, R. and others},
  booktitle={Proc. ISC High Perform.},
  pages={1--13},
  year={2025}
}

@article{furmento2025optimizing,
  title={Optimizing parallel heterogeneous system efficiency: Dynamic task graph adaptation with recursive tasks},
  author={Furmento, N. and others},
  journal={J. Parallel Distrib. Comput.},
  volume={205},
  pages={105157},
  year={2025}
}

@article{faverge2023programming,
  title={Programming heterogeneous architectures using hierarchical tasks},
  author={Faverge, M. and others},
  journal={Concurrency Computat. Pract. Exper.},
  volume={35},
  number={25},
  pages={e7811},
  year={2023}
}

@article{amestoy2023mixed,
  title={Mixed precision low-rank approximations and their application to block low-rank {LU} factorization},
  author={Amestoy, P. R. and others},
  journal={IMA J. Numer. Anal.},
  year={2023}
}

@article{charara2019batched,
  title={Batched Triangular Dense Linear Algebra Kernels for Very Small Matrix Sizes on {GPUs}},
  author={Charara, A. and Keyes, D. and Ltaief, H.},
  journal={ACM Trans. Math. Softw.},
  year={2019}
}

\end{document}